\documentclass[a4paper]{jpconf}
\usepackage{graphicx}
\usepackage{float}
\usepackage{wrapfig}

\usepackage{bm}
\usepackage{citesort}
\usepackage{threeparttable}
\usepackage[dvipsnames]{xcolor}

\usepackage[normalem]{ulem}  

\newcommand{\HI}{H\,{\sc i}}

\newcommand{\HeII}{He\,{\sc ii}}
\usepackage[textwidth=1in]{todonotes}

\begin{document}
\title{Equation of state of the intergalactic medium in the early Universe}

\author{K N Telikova$^{1,2}$, S A Balashev$^2$ and
P S Shternin$^2$}

\address{$^1$ Peter the Great St.\ Petersburg Polytechnic University,  29 Politeknicheskaya st., St.\ Petersburg, 195251, Russia}
\address{$^2$ Ioffe Institute, 26 Politeknicheskaya st., St.\ Petersburg, 194021, Russia}

\ead{ks.telikova@mail.ru}

\begin{abstract}
Spectroscopy of the Ly$\alpha$ forest in quasar spectra proved to be a useful tool for probing the intergalactic gas. We developed the automatic program for Voigt profile fitting of Ly$\alpha$ forest lines. We run this code on 9 high resolution ($\sim 50000$) quasars spectra with a high signal-to-noise ratio ($\sim 50 -100$) from the Keck telescope archive and obtained the sample of single well-fitted Ly$\alpha$ lines. Fitting the joint 2d distribution of column density and Doppler parameter from this sample by physically reasonable model we estimate a power law index, $\gamma$, of the equation of state of the intergalactic medium in the redshift range $z\sim 2-3$. 
We found that our measurement is in an agreement with measurements by other groups obtained with Voigt profile fitting technique.
\end{abstract}

\section{Introduction}
It is well known that the intergalactic medium (IGM) contains a dominant fraction of baryons in the Universe \cite{Meiksin2009}. After the recombination epoch at a redshift $z\sim 1100$ the intergalactic medium (IGM) had remained completely neutral until first stars, galaxies and quasars began to produce the ionizing background radiation. The IGM at this epoch mostly consisted from hydrogen and helium and consequently the IGM reionization included two major stages. The first stage is the reionization of \HI\ which was completed at $z > 6$ \cite{McGreer2015} and dominated by starlight escaped from the galaxies \cite{Robertson2015}. At this epoch the helium was already singly ionized. The second stage is the reionization of \HeII\ which occurs at later epochs, $z < 3$ \cite{Worseck2011}, and is driven by quasars radiation \cite{Madau2015}. Despite the considerable progress, our understanding of the reionization process is incomplete.
For instance, we still poorly know the properties of the ionizing ultraviolet background, precise redshifts of the reionization events and their impact on the thermal state of IGM. 
It is assumed \cite{HuiGnedin} that the equation of state -- the temperature-density, $T-\rho$, relation --  after the \HI\ reionization epoch is set by the competing processes of photo-heating and adiabatic cooling and obeys a power-law form
\begin{equation}\label{eos}
  T=T_0 \left( \frac{\rho}{\bar{\rho}} \right) ^{\gamma-1}\equiv T_0 \Delta^{\gamma-1},
\end{equation}
where $T_0$ is the temperature at the mean density $\bar{\rho}$, $\Delta$ is the overdensity and according to theoretical predictions. Numerical calculations suggest the index $\gamma$ is about 1.6 well after the reionization \cite{HuiGnedin,Puchwein2018}. 

\begin{figure}[t]
\centering
\includegraphics [width=1.0\textwidth]{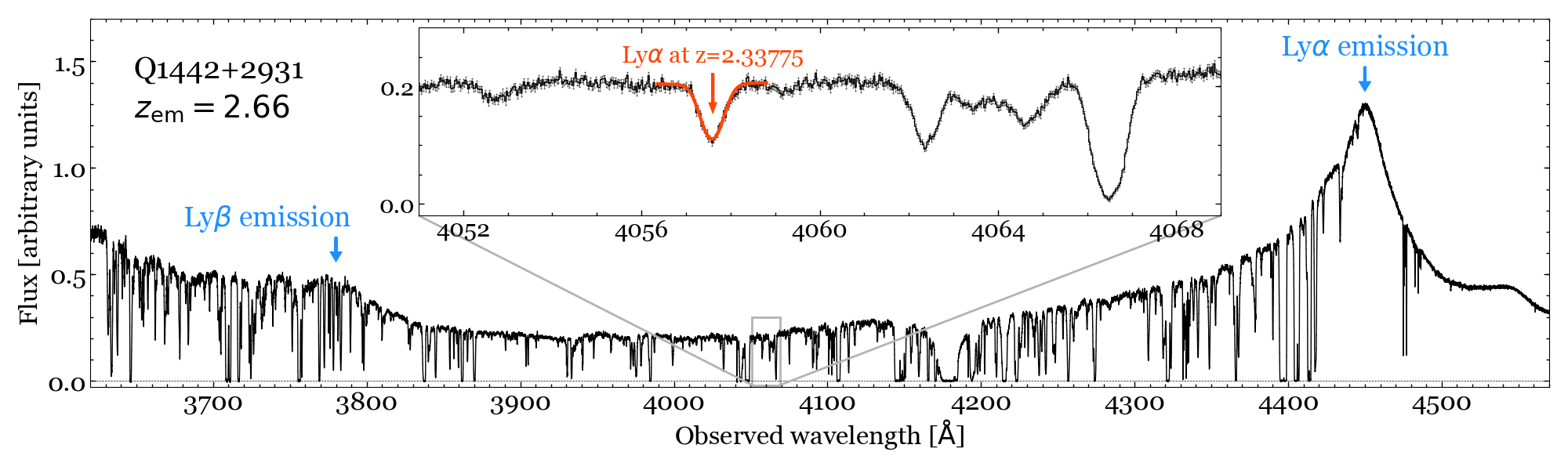}
\caption{The high resolution spectrum of the Q1442$+$2931 obtained with Keck telescope. The Ly$\alpha$ forest is the region bluewards the Ly$\alpha$ emission line of quasar. It consists of numerous Ly series \HI\ absorption lines associated with IGM. {\sl Inlay}: zoom of a small part of the spectrum; the red curve shows the fit to the \HI\ Ly$\alpha$ absorption line at $z\approx2.33775$.} 
\label{forest}
\end{figure}

The small residual neutral fraction of hydrogen remained after the reionization (at $z<6$) leaves traces in the spectra of background quasars observed as series of the absorption features called the Ly$\alpha$ forest (figure~\ref{forest}). An analysis of these lines allows 
to probe the thermal state of the IGM and its evolution with redshift. Various methods were suggested based on the statistical properties of the Ly$\alpha$ forest. These include analysis of the flux probability distribution function \cite{Viel2009,Lee2015}, the power-spectrum of the transmitted flux \cite{Zaldarriaga2001,Rorai2017a},
the average local curvature \cite{Becker2011,Boera2014} and wavelet decomposition of the Ly$\alpha$ forest \cite{Lidz2010,Garzilli2012}. In this paper we employ another widely-used  technique which is based on a decomposition of the Ly$\alpha$ forest into individual components and their subsequent Voigt profile analysis \cite{Schaye1999,Schaye2000,Ricotti2000,Rudie2012,Hiss2017,Gaikward2017,Rorai2018}.  

\section{Data and analysis}

Table~\ref{qsos} collects basic parameters of 9 selected high-resolution quasar spectra with high signal-to-noise ratio (SNR) we used for the analysis. Most of them was taken from KODIAQ database \cite{OMeara2017} and one spectrum, Q1442+2931, was reduced by hand from the Keck telescope archival data using \textsf{MAKEE} package\footnote{{\tt http://www.astro.caltech.edu/\~{}tb/makee/}}. An unabsorbed continuum level for the Q1442$+$2931 spectrum was estimated using B-spline interpolation, spectra from KODIAQ database are already normalized, but in some cases we adjusted continuum manually.

\begin{center}
\begin{table}[!t]
\centering
\caption{\label{qsos} Quasar spectra used for the analysis.}
\begin{tabular}{lccll} 
 \mr
 Quasar &$z_{\rm em}^{\dagger}$&$z_{\rm range}^{\ddagger}$&SNR&Resolution\\
 \mr
 J010311$+$131671&2.72&$2.06-2.68$&107&36000\\
 J014516$-$094517&2.72&$2.03-2.70$&83&36000\\
 J020950$-$000506&2.83&$2.15-2.79$&82&48000\\
 J082619$+$314848&3.09&$2.44-2.58$&44&48000\\
 J101155$+$294141&2.65&$2.11-2.60$&63&36000\\
 Q1442$+$2931&2.66&$2.08-2.62$&117&48000\\
 J155152$+$191104&2.84&$2.09-2.80$&96&36000\\
 J162548$+$264658&2.52&$2.11-2.50$&43&36000\\
 J170100$+$641209&2.73&$2.06-2.71$&70&36000\\
\br
\end{tabular}
\begin{tablenotes}
\item $\dagger$ quasar redshift 
\item $\ddagger$ redshift range used for analysis of the Ly$\alpha$ forest.
\end{tablenotes}
\end{table}
\end{center}

We developed an automatic procedure for analysis of the \HI\ absorption lines in the Ly$\alpha$ forest. In each quasar spectrum we used the spectral region between Ly$\alpha$ and Ly$\beta$ emission lines with offset of 12000 km/s bluewards from Ly$\alpha$ emission lines. This allowed us to avoid confusion with Ly$\beta$ forest lines and to minimize proximity effects for quasar. Each individual \HI\ absorber in the Ly$\alpha$ forest was described by the Voigt profile, with the column density, $N$, Doppler parameter, $b$, and redshift, $z$. Instead of fitting each Ly$\alpha$ forest line in specified redshift range with multicomponent profile, we searched for the solitary lines that can be fitted with only one component (see example in the inlay of figure~\ref{forest}). We constructed the dense grid of parameters ($z$, $N$, $b$), see table~\ref{grid}, and used $\chi^2$ criterion to compare model profiles calculated on that grid with observed ones. The Ly$\alpha$ forest lines in the probed redshift range, $z \sim 2-3$, frequently blend with each other. Therefore, we used a sort of outlier rejection routine to include in analysis also the lines with weak blends in one of the wings. 
In addition we checked that the higher-order Lyman-series absorption lines are in agreement with the estimated parameters from corresponding Ly$\alpha$ line. Finally, we excluded lines with parameters lying on the grid edges to get rid of the boundary effects on the line sample density estimation in the parameters space.
In this way we obtained the representative sample of "solitary" Ly$\alpha$ forest lines in the specified range of ($N$, $b$) parameters. 

\begin{figure}[!tbp]
  \centering
  \begin{minipage}[b]{0.48\textwidth}
    \includegraphics[width=\textwidth]{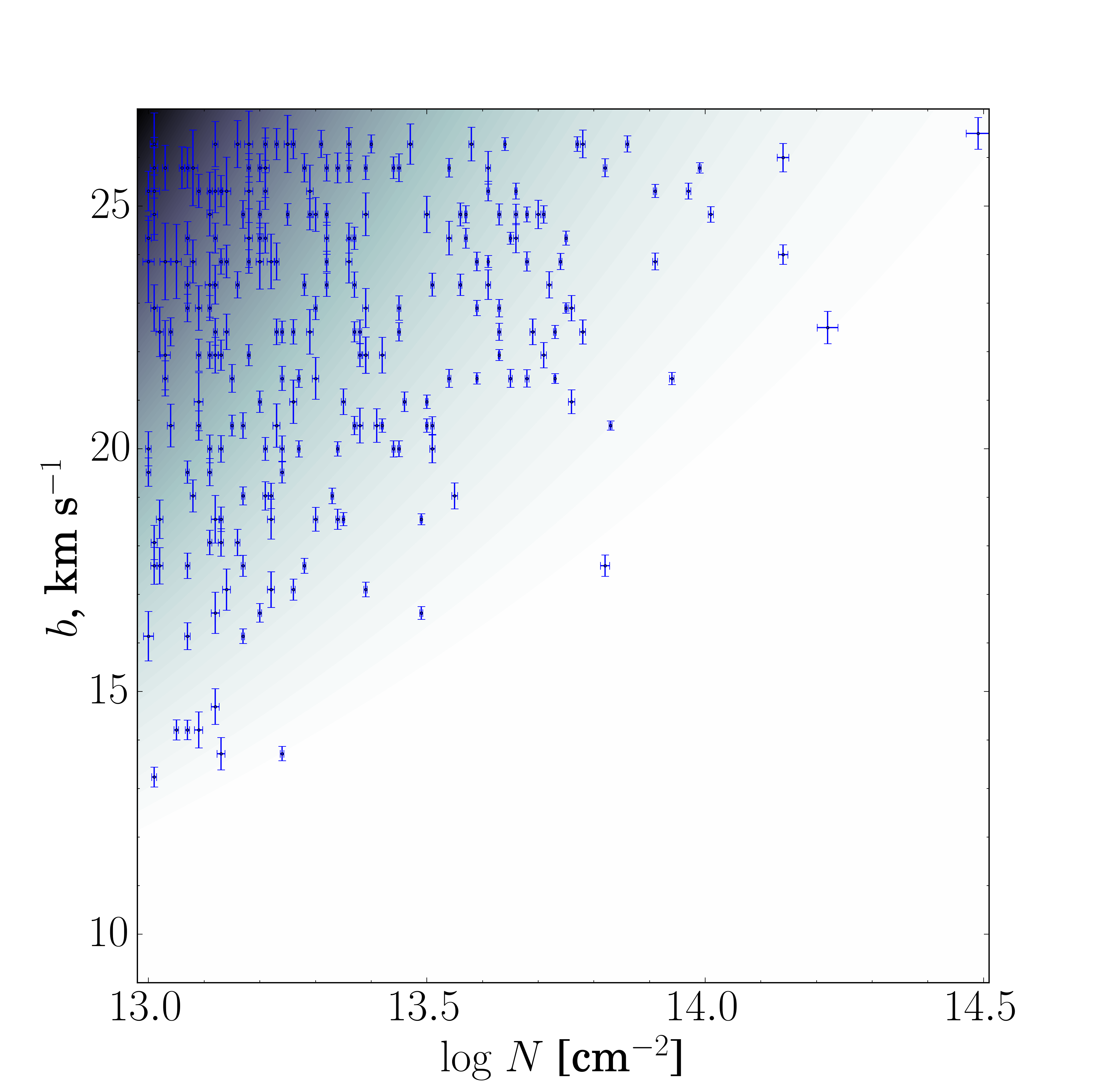}
    \caption{Error crosses show the obtained absorption lines sample. The gradient-filled area shows the best-fit probability density function of the ($N,b$) sample.}\label{fig:data}
  \end{minipage}
  \hskip0.05\textwidth
  \begin{minipage}[b]{0.45\textwidth}
    \includegraphics[width=\textwidth]{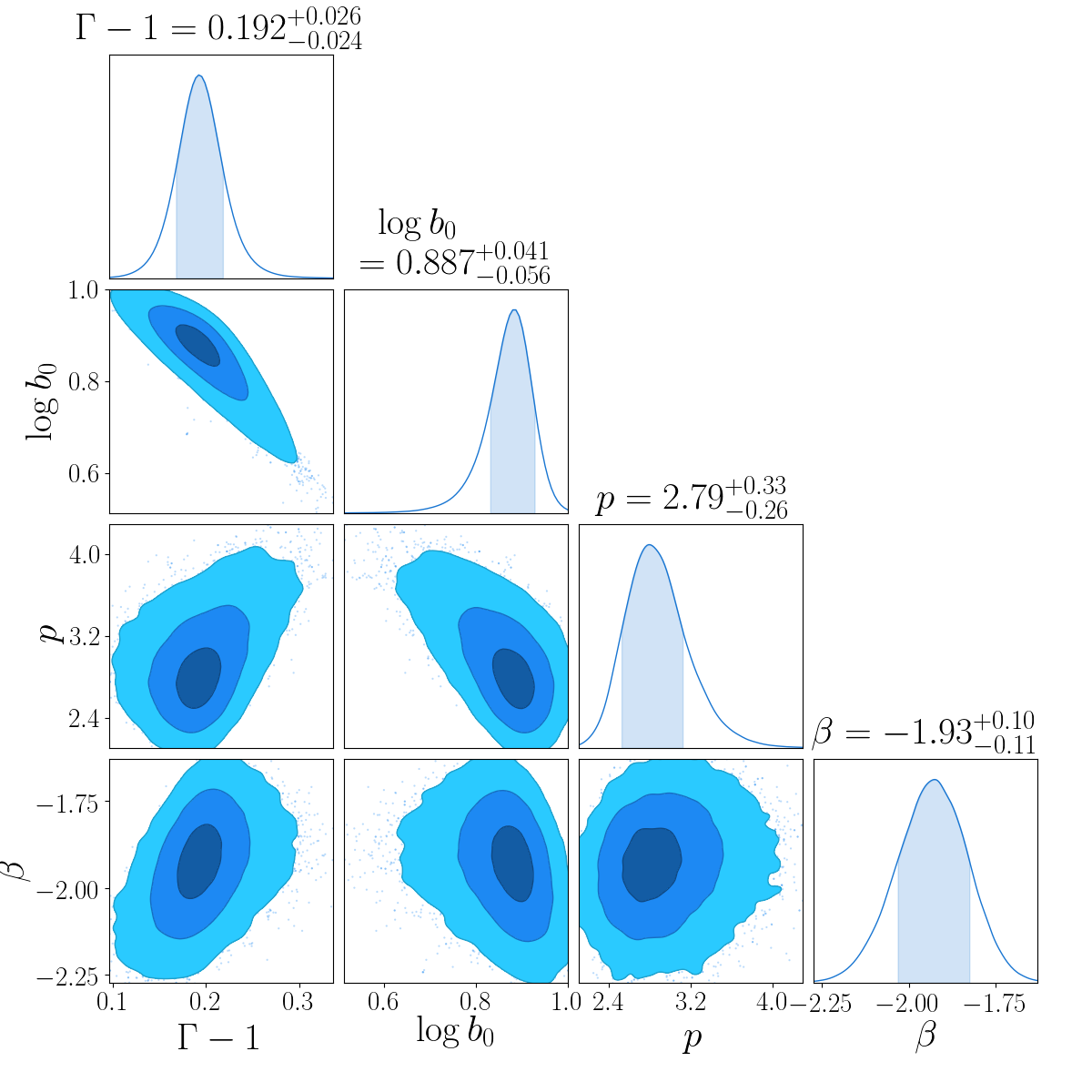}
    \caption{Marginal posterior distributions of the parameters $\Gamma$, $\log b_0$\,[km\,s$^{-1}$], $p$ and $\beta$. Dark and light blue areas shows 1$\sigma$ and 2$\sigma$ levels for 1D distributions respectively.}\label{fig:mcmc}
  \end{minipage}
\end{figure}

Since our selection procedure is not the line profile fitting, it does not return the line parameter uncertainties  $\sigma^N$ and $\sigma^b$. These were obtained analytically via the Fisher matrix for the $\chi^2$ likelihood function with account for resolution and SNR of each spectra. We checked that this gives reasonably close estimates of the expected uncertainties would the real fit be performed. The resulting sample in ($N, \,b$) parameter space is shown in figure~\ref{fig:data}. The mean redshift of the Ly$\alpha$ lines in our sample is $\bar{z}=2.35$. 




\begin{center}
\begin{table}[t]
\centering
\caption{\label{grid}Parameters of the grid.}
\begin{tabular}{@{}l*{15}{l}}
\br
Parameter&Range&Step\\
\mr
Redshift, $z$&Individual&$8.12 \times 10^{-6}$\\
Column density, $\log N~[\rm{cm}^{-2}]$&$13-15$&0.02\\
Doppler parameter, $b$ [$\rm{km} \ \rm{s}^{-1}] $&$12-27$&0.48\\
\br
\end{tabular}
\begin{tablenotes}
\item The choice of steps in the grid was based on the Fisher matrix estimate.
\end{tablenotes}
\end{table}
\end{center}

The obtained $(N,\,b)$ distribution shows a prominent lower envelope. Its physical origin is as follows \cite{Schaye1999}. The parameter $b$ characterizes the width of the velocity distribution of hydrogen atoms, which contains thermal and turbulent motions inside each absorber.
Since thermal broadening is unavoidable, for each $N$ there exists a minimal value of $b$ which corresponds to a pure thermal broadening
\begin{equation}\label{minb}
  b\to b_{\rm{th}} \equiv \sqrt{2k_{\rm B}T/m},
\end{equation}
where $k_{\rm B}$ is the Boltzmann constant and $m$ is the hydrogen atomic mass.


The position of the low boundary of the $N-b$ distribution can be connected to the temperature-density relation of the IGM, given by eq~(\ref{eos}). 
In assumption of uniform background radiation, the \HI\ column density $N$ is related to the overdensity $\Delta$ as \cite{Schaye2001,Bolton2014}\footnote{The value of the prefactor in eq~(\ref{rho-N}) and the power exponent of $T$ there depend on the adopted value of case-A hydrogen recombination coefficient. Here we take the same prescription as in \cite{Bolton2014}.}
\begin{equation}\label{rho-N}
	N = 1.3\times 10^{14}\, \Delta^{3/2}\frac{T^{-0.22}}{\Gamma_{-12}}\left(\frac{1+z}{3.4}\right)^{9/2}~{\rm cm}^{-2},
\end{equation}
where $\Gamma_{-12}$ is the hydrogen photoionization rate in units of $10^{-12}$~$\rm s^{-1}$. 
Eqs~(\ref{eos}), (\ref{minb}) and (\ref{rho-N}) allow to predict the $b_{\rm th}-N$ dependence
\begin{equation}\label{gamma-Gamma}
 b_{\rm{th}}= b_0\left(\frac{N}{10^{12}~{\rm cm}^{-2}}\right)^{\Gamma-1}, \quad \Gamma-1 = \frac{\gamma-1}{3-0.44(\gamma-1)},
\end{equation}
where the normalization constant $b_0$ depends on $T_0$, $\Gamma_{-12}$ and $z$. Once $b_0$ and $\Gamma$ are found, the parameters  $\gamma(z)$ and $T_0 (\Gamma_{-12},\,z)$ can be inferred.


Techniques that employ decomposition of Ly$\alpha$ forest lines are usually based on direct determination of the lower boundary of $N-b$ distribution. This is not a straightforward task. Various iterative techniques are proposed to eliminate the turbulent contribution to width of the lines \cite{Schaye2000,Rudie2012,Hiss2017}. Most popular statistics is based on the so-called 2$\sigma$ rejection algorithm \cite{Schaye1999}, which in some cases is calibrated with the hydrodynamical simulations.

Here we propose to fit the whole $(N,\,b)$ distribution.
We use microturbulence assumption that thermal and turbulent motions are uncorrelated and can be characterized by the Gaussian distributions with widths $b_{\rm th}$ and $b_{\rm turb}$, respectively. The convolution of these distributions is also Gaussian with the characteristic width $b = \sqrt{b_{\rm{th}}^2 + b_{\rm{turb}}^2}$.
We assume that turbulent motions are independent on the thermal state of the gas and its density. Besides, we suggest that distributions over $N$ and $b_{\rm turb}$ have power-law shapes with indexes $\beta$ and $p$, respectively. Notice that for $N$ distribution this is a standard assumption and index $\beta$ is well measured, see, e.g., \cite{Ricotti2000}. In this way the 2D probability density distribution has the form
\begin{equation}\label{pdf}
  f(N, b)=\int f_N(N)f_{\rm turb}(b_{\rm{turb}})\delta \left(b-\sqrt{b_{\rm{th}}^2+b_{\rm{turb}}^2} \right)\rm{d} \emph{b}_{\rm{turb}},
\end{equation}
where $f_{N}$ and $f_{\rm turb}$ are the probability density distributions over $N$ and $b_{\rm turb}$, respectively. Thus the function $f$ depends on the parameter vector  $\bm{\Theta}$=($\Gamma$,\, $\log b_0$,\, $p$,\, $\beta$) in addition to $N$ and $b$.
To estimate the posterior distributions of the parameters $\bm{\Theta}$ we used the Bayesian framework. Flat priors were chosen for all parameters.
Our $(N,\, b)$ sample probes only a limited part of the full distribution $f(N,b)$, namely the region specified by the searching procedure box (see table~\ref{grid}). The estimated uncertainties $\sigma^N_{i}$ and $\sigma^b_{i}$ in the sample are heteroscedastic, therefore the probability of the data point to fall in this box is different for each line, since this probability depends on the observed $(N,b)$ values and not on the unknown true values $(\widetilde{N},\widetilde{b})$.
As a result, the correctly normalized likelihood function is \cite{Kelly2007}
\begin{equation}\label{Likelihood}
{\cal L}\left(N_i,\, b_i |\,\bm{\Theta}\right) =\prod_i \frac{\int f(\widetilde{N}, \widetilde{b})f_i(N_{i},b_{i}|\,\widetilde{N},\widetilde{b})\,{\rm d}\widetilde{N}\,{\rm d}\ \widetilde{b}}
{\int f(\widetilde{N}, \widetilde{b})I(N, b)f_i(N,b|\,\widetilde{N}, \widetilde{b})\,{\rm d}N\,{\rm d}b\,{\rm d}\widetilde{N}\,{\rm d}\widetilde{b}},
\end{equation}
where $f_i(N,b|\,\widetilde{N},\widetilde{b})$ is the bivariate normal distribution of parameters $N,b$ with the mean $\widetilde{N},\, \widetilde{b}$ and the dispersion $\sigma_i^{N},\, \sigma_i^{b}$. $I(N,\,b)$ is the selection function equal to one if a pair $(N,\,b)$ falls in our analysis box and zero otherwise.
We performed the maximum likelihood estimate of  the distribution parameters $\bm{\Theta}$ employing the Markov Chain Monte Carlo (MCMC) simulations with the affine sampler \textsf{emcee} \cite{Foreman-Mackey2013}.

\section{Results and discussion}

The marginal posterior distributions of the fit parameters are shown in figure~\ref{fig:mcmc}. According to eq~(\ref{gamma-Gamma}), the estimated power-law index in the temperature-density relation~(\ref{eos}) of the IGM at the average redshift $z=2.35$  is $\gamma-1 = 0.53 \pm 0.07$ at the 68 per cent confidence level. Using eq~(\ref{rho-N}) we find the temperature at the mean density
\begin{equation}\label{T0-infer}
\log T_0\, [K] = (3.98\pm 0.05) - (0.35\pm 0.04) \log \Gamma_{-12}
\end{equation}
which, however, depends on the photoionization rate $\Gamma_{-12}$. 
From the analysis of the mean opacity of the Ly$\alpha$ forest lines it was constrained that $\left(T_0 [K]\right)^{-0.72}/\Gamma_{-12}=(1.88\pm0.08)\times 10^{-3}$ at $z=2.4$ \cite{Faucher2008ApJ}. 
Their calculations also depend on the equation of state of the IGM, however, their adopted value for $\gamma-1=0.6$ is rather close to our $0.53$ (see their figure 3), so we do not expect a significant bias. Adopting the result of \cite{Faucher2008ApJ} we can constrain both 
$T_0=1.25^{+0.05}_{-0.19}\times 10^4$~K and $\Gamma_{-12}=0.60^{+0.08}_{-0.03}$ with a strong anticorrelation between these parameters. 

The sample is much smaller than some samples derived from the multicomponent fit (see e.g.~\cite{Rudie2012,Hiss2017}), but it allows us to avoid a lack of knowledge of ``true" multicomponent models and correlations of the parameters of such compound models. This can be comprehended if one compares our sample with sample, obtained on the same quasar spectra \cite{Rudie2012}: at each column density region the Rudie et al detect much more lines with low $b$ values than we did. But finally they rejected most of the lines with low $b$ values by some arbitrary routine.

In previous sections we suggest that broadening of absorption lines determined by  thermal and turbulent motions. Turbulent broadening is driven by peculiar velocities of the absorbed intergalactic clouds and in our approach it is independent on column density $N$. But in general real turbulent mechanisms may depend on $N$. For example we did not take into account expected dependence of $b$ on $N$ due to pressure smoothing scale and differential Hubble flow along to physical extent of intergalactic clouds \cite{Gnedin&Hui1998,Kulkarni2015,Garzilli2015}.

\section{Summary}
We developed the automatic procedure for searching and fitting Ly$\alpha$ forest absorption lines and applied it to 9 high-resolution quasar spectra with high SNR in the redshift range $z \sim 2-3$. Analysis of the full distribution of the absorption line parameters allowed us to probe the equation of state of the IGM at $\bar{z}=2.35$. The power-law index of the temperature--overdensity relation $\gamma-1 = 0.53 \pm 0.07$ is in a broad agreement with the results of other authors. We also obtained the relation (\ref{T0-infer}) between the temperature $T_0$ at the mean density $\bar{\rho}$ and the hydrogen photoionization rate $\Gamma_{-12}$.  Combining this result with measurements of \cite{Faucher2008ApJ} we were able to separately constrain  $T_0$ and $\Gamma_{-12}$. The proposed automatic program will allow us to analyze in the future much more quasar spectra available in Keck and VLT archives. Such increase of statistic is strongly important to precisely estimate the redshift evolution of the main parameters described the IGM equation of state -- $\gamma$, $T_0$ and $\Gamma_{-12}$.

\section*{References}
\bibliographystyle{iopart-num}
\bibliography{references.bib}
\end{document}